\documentclass[runningheads]{llncs}
\usepackage{graphicx}
\usepackage{times}
\usepackage{enumerate}   
\usepackage{booktabs}
\usepackage{graphicx}
\usepackage[export]{adjustbox}
\usepackage[breaklinks]{hyperref}
\usepackage{csquotes}
\usepackage[capitalise,nameinlink]{cleveref}
\crefname{section}{Section}{Sections}
\Crefname{section}{Section}{Sections}
\crefname{figure}{Figure}{Figures}
\Crefname{figure}{Figure}{Figures}
\crefname{listing}{\lstlistingname}{\lstlistingname}
\Crefname{listing}{Listing}{Listings}
\usepackage{draftwatermark}
\SetWatermarkText{preprint}
\SetWatermarkScale{0.7}

\usepackage{url}

\usepackage{breakurl}

\usepackage{caption}
\begin{document}
\title{Pattern Views: \\ Concept and Tooling for Interconnected \\ Pattern Languages}
%
%
\author{Manuela Weigold\orcidID{0000-0002-4554-260X} \and
Johanna Barzen\orcidID{0000-0001-8397-7973} \and
\\Uwe Breitenbücher\orcidID{0000-0002-8816-5541} \and
Michael Falkenthal\orcidID{0000-0001-7802-1395} \and
\\Frank Leymann\orcidID{0000-0002-9123-259X} \and
Karoline Wild\orcidID{0000-0001-7803-6386}}
\authorrunning{Weigold et al.}
%
\institute{
Institute of Architecture of Application Systems, 
\\University of Stuttgart, Universitätsstrasse 38, Stuttgart, Germany \\
\email{\{firstname.lastname\}@iaas.uni-stuttgart.de}}
\maketitle              
\begin{abstract}
Patterns describe proven solutions for recurring problems.
Typically, patterns in a particular domain are interrelated and organized in pattern languages.
As real-world problems often require patterns of multiple domains, different pattern languages have to be considered to address these problems.
However, cross-domain knowledge about how patterns of different languages relate to each other is either hidden in individual pattern descriptions or not documented at all.
This makes it difficult to identify relevant patterns across pattern languages.
Therefore, we introduce a concept and tooling that enables to capture  patterns and their relations across pattern languages for a particular problem context.

\end{abstract}
\section{Introduction}
Patterns describe proven solutions for recurring problems. 
After the first patterns were published in the domain of city and building architecture by Alexander~et~al.~\cite{Alexander1977_PatternLanguage}, the concept of patterns has been adopted in various other fields.
Especially in software and information technology, publishing patterns has become a popular way to convey expert knowledge in different domains, e.g., object-oriented programming~\cite{Gamma1995_PatternLanguages}, enterprise application architecture~\cite{Fowler2002_PatternLanguages}, messaging~\cite{Hohpe2003_PatternLanguages}, or security~\cite{schumacher2013security}. 
Since patterns can often be used in combination or offer alternative solutions, the relations between patterns are essential for identifying all relevant patterns and are therefore also documented. 
For example, the cloud computing pattern \emph{Public Cloud}~\cite{Fehling2014_Book} describes how cloud providers can offer IT resources to a large customer group. 
It further refers to patterns that describe the different service models for offering resources, e.g., as \emph{Infrastructure as a Service (IaaS)}. 
When using the Public Cloud pattern, those related patterns should also be considered.
In conjunction with the relations between them, patterns can be organized in pattern languages~\cite{Alexander1977_PatternLanguage}. 
As a result, a pattern language describes how patterns work together to solve broader problems in a particular domain~\cite{Coplien1996_SoftwarePatterns}.

\newpage
However, real-world problems often require patterns of
different domains. Most likely, not all relevant patterns belong
to the same pattern language. Therefore, some authors include
references to other languages, e.g., Fehling et al. [6] state that
the message queues of the \emph{Message-oriented Middleware} pattern in the cloud computing patterns are introduced by Hophe~\&~Woolf's~\cite{Hohpe2003_PatternLanguages} pattern language as \emph{Message Channels}. Unfortunately, not all relevant pattern languages are referred. For example, distributed
cloud applications typically have to meet security requirements
regarding the communication of the distributed components. To
ensure secure communication, Schumacher et al.’s~\cite{schumacher2013security} \emph{Secure Channel} pattern can be applied. However, this pattern language
is not mentioned by Fehling et al.~\cite{Fehling2014_Book} at all. As references
to other pattern languages are often missing, it is difficult to
identify relevant patterns in other areas.

One reason for missing references is the way pattern languages are documented. 
Most pattern languages are published in books or scientific publications. 
Once they are published, they can hardly be changed and therefore the patterns languages remain static.
This was not intended by Alexander~et~al.~\cite{Alexander1977_PatternLanguage}, who described them as \emph{living networks}.
Some authors created dedicated websites for their pattern languages (e.g., \cite{cloudComputingPatterns_website,internetofthingspatterns_website,enterpriseintegrationpatterns_website}), which eases the adaptation. 
Nevertheless, these websites represent only one particular language. 
For this reason, pattern repositories have been developed that aim to collaboratively collect patterns of various domains and provide tooling support to edit or extend patterns and relations. 

Although several pattern repositories support the collections of patterns, patterns from different languages are sometimes treated only as a simple interconnected set and are not organized in pattern languages (e.g.,~\cite{uipatterns_repo,designpatterns_catalog}).
However, a pattern language is more than a collection of patterns. 
It reflects the higher-level \emph{domain} for which the patterns are relevant~\cite{borchers_pattern_2008}, e.g., for realizing cloud applications. 
A few repositories organize patterns in pattern languages (e.g.,~\cite{publicSphereProject_repo,learningenvironmentslab_repo}), but do not explicitly reflect cross-domain relations between patterns in different languages.
This {knowledge} is hidden in individual pattern descriptions.
However, without explicitly cross-domain relations, and without the context in which these relations are relevant, it is difficult to identify relevant patterns for a given problem.
This means that it must be possible (i) to assign patterns to a particular pattern language, (ii) to document cross-domain relations, and (iii) to specify the context in which a set of patterns and their relations are relevant.

To tackle these issues, we introduce a concept to explicitly document cross-domain knowledge relevant for a particular problem context.
For this, patterns and their relations from different pattern languages can be selected and further relations can be defined as relevant in a specific context.
The relations between patterns of different languages are \emph{cross-language relations} that express cross-domain knowledge.
Thus, it is possible to combine and extend pattern languages -- a truly \emph{living network of patterns}.
Based on our previous experience with pattern repositories, we show how support for the concept can be integrated into the tool chain of an existing pattern repository. 
We also present a prototype that supports multiple pattern languages and their dynamical combination by using the new concept. 
The remainder of the paper is structured as follows: \cref{sec:motivation} describes  fundamentals and a motivating scenario, \cref{sec:patternatlas} introduces our concept and tooling for it. 
We further present a use case in \cref{sec:use-case}. 
Finally, \cref{sec:related-work} describes related work and \cref{sec:conclusion} concludes the paper.

\begin{figure*}[t]
    \centering
     \includegraphics[width=\textwidth]{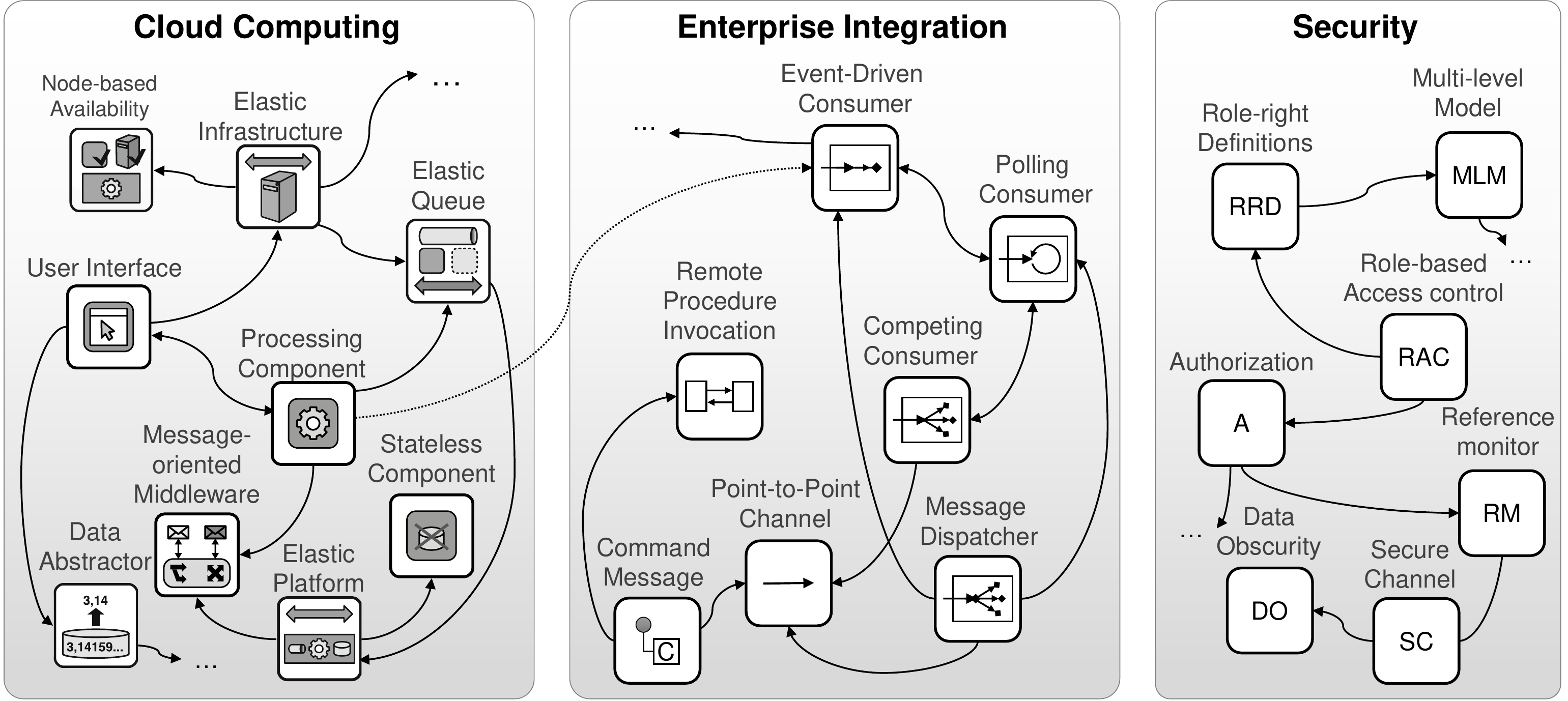}
    \caption{Patterns and relations of multiple pattern languages: Cloud computing patterns~\cite{Fehling2014_Book}, enterprise integration patterns~\cite{Hohpe2003_PatternLanguages}, and security patterns~\cite{schumacher2013security}.}
    \label{fig:motivation}
\end{figure*}

\section{Background and Motivation}
\label{sec:motivation}
In this section, we first introduce patterns and pattern languages and then motivate that for real-world problems often patterns from multiple domains have to be considered.

\subsection{Patterns and Pattern Languages} 
\label{subsec:PatternLanguages}
As already mentioned, patterns are used to gather knowledge about proven solutions for recurring problems in many different fields, e.g., humanities~\cite{barzen_pattern_humanities} or software engineering~\cite{Coplien1996_SoftwarePatterns}.
They describe the core idea of the solution on an general manner, which means in case of software engineering patterns that they are independent of a specific technology or programming language.
The general solution concept of a pattern can therefore be applied to a variety use cases in different ways.
Since humans are the targets, patterns are documented as textual descriptions according to a defined pattern format. 
Even if the pattern formats differ slightly from pattern language to pattern language~\cite{henninger_software_2007}, typical formats for patterns in software and information technology domains contain sections about the addressed \emph{problem}, the \emph{context} in which the problem might arise, \emph{forces} which direct the problem, the proposed \emph{solution}, the \emph{resulting context} describing which forces have been resolved, and a section showing a \emph{sketch} of the solution~\cite{coplien_software_1996}. 
Often other patterns are only referenced in the textual description of one of these sections.
Some authors have introduced explicit sections to describe the relations of the pattern and give them defined semantics~\cite{Falkenthal2018_NatureOfPatternLanguages}, such as \emph{Variations}~\cite{Fehling2014_Book,internetofthingspatterns_website}, \emph{See also}~\cite{schumacher2013security} or \emph{Next}~\cite{Hohpe2003_PatternLanguages}. 

Patterns and relations are the basic building blocks of pattern languages. 
In this work, we build on the premise that a pattern language is more than a collection of~patterns, but a designed system~\cite{winn2003pattern}. 
This means that (i) relations of a pattern language are designed to guide the reader towards suitable patterns and (ii) each pattern solves a specific problem that is related to the overall context of the pattern language~\cite{Meszaros1997_PatternWriting,borchers_pattern_2008}, e.g., in the domain of cloud computing, enterprise integration, or security.

\subsection{Motivating Scenario}
\label{subsec:CrossLanguageRelations}
Often patterns of several domains have to be considered for a real-world problem. 
For example, suppose a software developer wants to build a secure elastic cloud application.
An elastic application responds to changing workload by adjusting the amount of resources allocated to the application~\cite{Fehling2014_Book}. 
This ensures that neither too many resources (which is costly) nor too few resources are allocated over a long period. 

The cloud computing patterns in \cref{fig:motivation} on the left provide several patterns relevant for an elastic cloud application: For example, an \emph{Elastic Infrastructure} provides a dynamically adjustable infrastructure to a customer to deploy an application and an \emph{Elastic Queue} can be used to monitor the number of message in a queue and to adjust the number of processing components handling the requests. 
In context of an elastic cloud application, the \emph{Processing Components} are often implemented as \emph{Competing Consumers} as any of the instances can receive and process an incoming request. Therefore, this enterprise integration pattern is explicitly referred to in the processing component pattern.
Since messaging is often used for integrating cloud application components, the cloud computing patterns contain several messaging-related patterns, such as the \emph{Message-oriented Middleware} pattern that also refer to other enterprise integration patterns.
However, often references to related pattern languages are missing. 
For example, the enterprise integration patterns were published before the cloud computing patterns and thus never reference them. 
And although most elastic cloud applications must meet certain security requirements, such as secure communication between application components as provided by the \emph{Secure Channel} pattern of the security patterns no security patterns are mentioned and thus no cross-language relations exist. 
It can easily be seen that cross-language relations are also important for pattern languages of other areas than software, e.g., for realizing a film scene, patterns from different domains (costumes, music and film settings) are needed~\cite{Falkenthal2015_PatternRefinement}.

But even if cross-language relations exist, they are often not properly documented.
The pattern languages depicted in \cref{fig:motivation} are published in books~\cite{Fehling2014_Book,Hohpe2003_PatternLanguages,schumacher2013security} or on dedicated websites~\cite{cloudComputingPatterns_website,enterpriseintegrationpatterns_website}. 
Besides scientific publications and dedicated websites, patterns are published in repositories that aim to collect patterns in collaboration~\cite{Fehling2014_Paper_PatternPedia}. 
However, even with the tooling support of current repositories it is challenging to find relevant patterns for a given problem: 
Several repositories do not organize patterns in pattern languages~\cite{uipatterns_repo,designpatterns_catalog} and treat patterns only as a simple interconnected set.
The few repositories organizing patterns in pattern languages~\cite{Fehling2014_Paper_PatternPedia,publicSphereProject_repo} hide cross-language relations in individual pattern descriptions. 
None of the repositories known to us enables to document patterns and relations for a specific context (e.g., secure elastic cloud application). 
Consequently, finding suitable patterns across pattern languages for a certain problem is a cumbersome, manual~process.

\begin{figure*}[t]
    \centering
     \includegraphics[width=\textwidth]{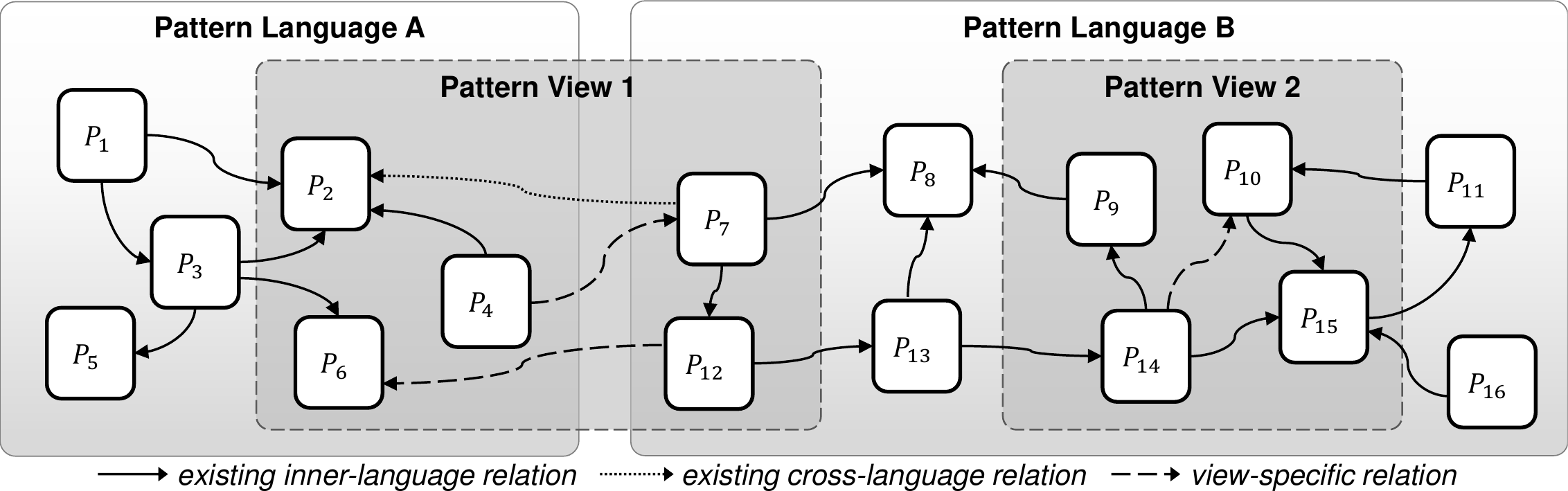}
    \caption{The concept of pattern views: Pattern views can comprise patterns of multiple (pattern view 1) or one pattern language (pattern view 2).}
    \label{fig:patternlanguages_to_View}
\end{figure*}

\section{Pattern Views}\label{sec:patternatlas}
In the following sections, we introduce \emph{pattern views} as a concept to document cross-domain knowledge for a particular context that requires patterns and relations across pattern languages.
We explain how pattern views can be integrated into the tool chain of a pattern repository by presenting our prototype. 

\subsection{The Concept of Pattern Views}
Already Alexander et al.~~\cite{Alexander1977_PatternLanguage} mentioned in the publication of the first pattern language that if a certain problem requires patterns of different pattern languages, patterns of different domains can be combined arbitrarily.
Based on this idea we introduce \emph{pattern views} as a concept (i) to explicitly define the context for which a set of patterns and relations is relevant, (ii) to specify new relations between patterns that are relevant in this specific context, and (iii) to preserve knowledge about the pattern languages from which the patterns originate.

\cref{fig:patternlanguages_to_View} illustrates our concept.
A \emph{pattern view} comprises patterns of either different pattern languages (pattern view 1) or one pattern language (pattern view 2).
For example, patterns from different languages are relevant for a secure elastic cloud application, while only a subset of the cloud computing patterns are relevant for the deployment of a cloud application. 
The relations between the contained patterns in a pattern view are either those already defined in the original language or newly defined relations that are relevant in the defined context of the pattern view.
Especially cross-language relations are often not embedded in the original languages.
The relevance of a pattern view is determined by its context.
The context guides the pattern users, e.g., software architects, to identify a pattern view for his or her particular problem.
Thus, pattern views enable to document knowledge about combining patterns and pattern language for a particular problem explicitly and reusable for other users.
In \cref{sec:use-case} a pattern view containing patterns relevant in the context of \emph{secure elastic cloud applications} is described in detail.
As a result, an individual pattern can be considered from different perspectives: It is primarily anchored in its original pattern language, but can also be part of different views that place the pattern in a specific context.
As a pattern view can reuse and extend the existing structure of underlying pattern languages, new structures emerge. This supports the notion of Alexander's \emph{living network of patterns} that is constantly~changing.  

The term \emph{pattern view} is inspired by two existing concepts in computer science: 
In database management systems, database views can be used to represent subsets of data contained in regular tables. 
They can join, aggregate, or simplify data from multiple tables and represent them as a single database view. 
For patterns, the same can be done by our pattern views: 
Patterns from multiple data sources (pattern languages) can be included in a pattern view.
New relations for the pattern view can be defined, just like a database view can refer to another table.
Another analogy to pattern views is the notion of architecture views in architecture descriptions~\cite{IEEE1471}. 
An architecture view represents the architecture of a system from a specific viewpoint that is in accordance with a certain set of stakeholders' concerns~\cite{IEEE1471}.
Depending on the concerns of the different stakeholders, a suitable architecture description can be created, e.g. a process view for process architects or a software distribution view for software developer. 
While Avgeriou~\&~Zdun~\cite{Avgeriou2005_Architecturalpatterns} use this definition to assign architectural patterns to their primary architectural view, e.g., the \emph{Client-Server} pattern to the component-interaction view, we only adopt the idea of views and define pattern views as a representation of pattern languages from a particular viewpoint.
The context of a pattern view represents the viewpoint from which the patterns and pattern languages are viewed to address the concerns of the pattern user.

In section \cref{sec:use-case} we present a pattern view for \emph{secure elastic cloud applications} that is aimed towards cloud software architects and contains several patterns for the integration of the application components. 
Even if we motivate our work on the basis of information technology pattern languages, our concept is not limited to them but can be applied, e.g., to patterns for film costumes~\cite{barzen_costumlanguages} or building architecture~\cite{Alexander1977_PatternLanguage}.

\subsection{Tooling for Pattern Views}
In previous works, \emph{PatternPedia}\footnote{https://github.com/PatternPedia}~\cite{Fehling2014_Paper_PatternPedia} has been introduced as a collaborative tool for documenting and managing patterns and pattern languages, as well as concrete solutions that are implementations of the patterns with a particular technology or in a particular programming language, in case of software engineering patterns.  
Pattern research is actively supported, as experts can analyze concrete solutions in collaboration and as a result identify best practices and document patterns~\cite{Fehling2014_Paper_PatternPedia}.

\begin{figure}[!t]
    \centering
    \includegraphics[width=\textwidth]{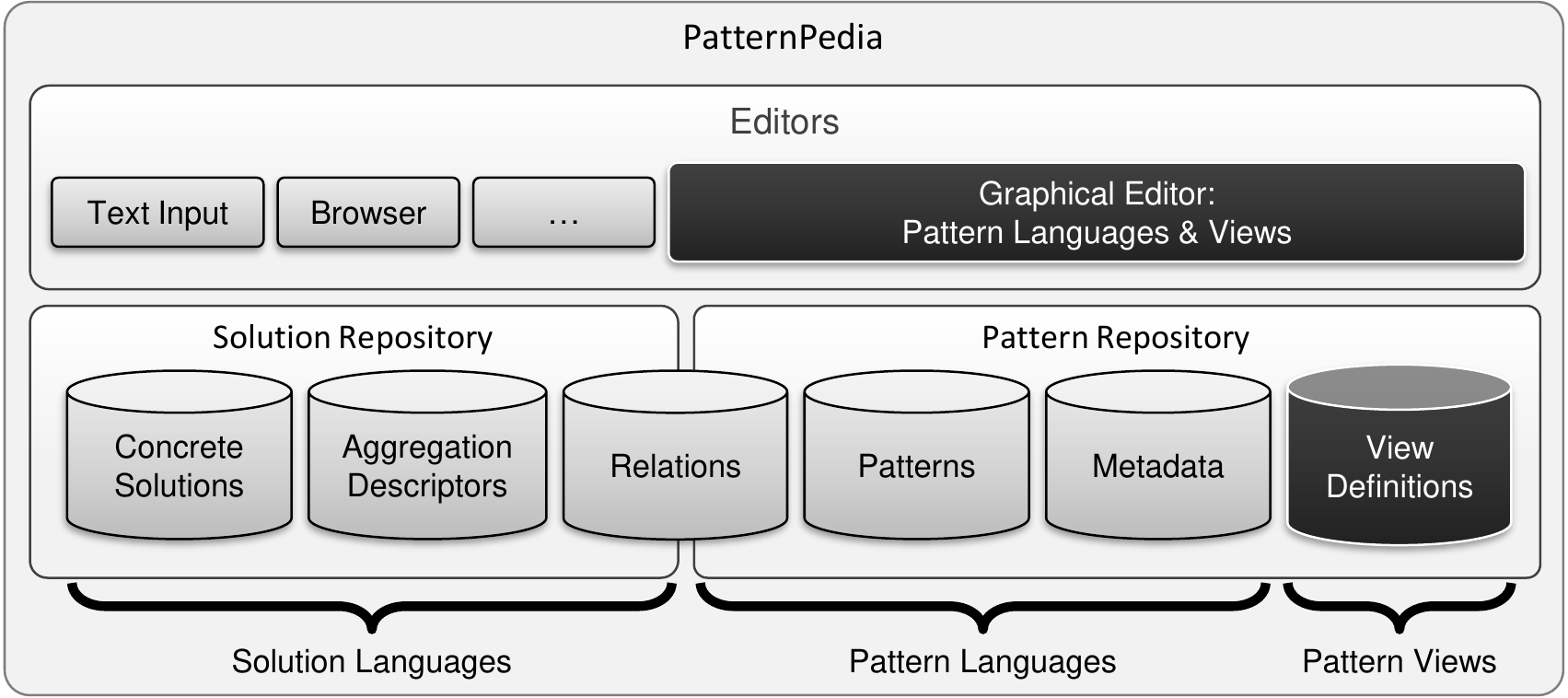}
    \caption{Abstract Architecture of PatternPedia.}
    \label{fig:patternpedia-extension}
\end{figure}

\cref{fig:patternpedia-extension} illustrates the abstract architecture of PatternPedia with the newly developed components in dark grey.
In the pattern repository patterns and relations between them are managed. 
The patterns as well as their relations are organized in pattern languages.
The metadata defines the pattern formats for the different pattern languages as well as the semantics of the relations. 
Analogously, the solution repository stores concrete solutions and their relations, which are organized in solution languages.
In addition, aggregation descriptors are stored that specify how different concrete solution artefacts can be combined~\cite{Falkenthal_2017_SolutionLanguages}.
The aggregation descriptors are used to annotate the relations between the concrete solutions in the solution languages.
The solution repository for managing solution languages depends highly on the domain of the solution, e.g., for concrete solutions of costumes detailed descriptions of clothing pieces are relevant~\cite{barzen_2018_muse} whereas solution of software patterns can be code snippets~\cite{Fehling2014_Paper_PatternPedia}.
The PatternPedia editor facilitates to add patterns as textual descriptions and browse the patterns and pattern languages as well as solution languages.

In this work, we enriched the pattern repository of PatternPedia by the concept of pattern views and a graphical editor\footnote{A demo can be found here: \url{https://bit.ly/391T7Wz}}. 
The repository has been extended by the functionality to create pattern view definitions that specify the context of a particular pattern view and reference the existing relations and patterns across pattern languages that are part of the pattern view. 
Additionally, view-specific relations can be defined in the context of a particular pattern view.
These view-specific relations add additional knowledge about how patterns in a certain context are interrelated.

As pattern and their relations are commonly represented as a directed graph with nodes representing patterns and edges representing relations~\cite{Falkenthal2018_NatureOfPatternLanguages,Coplien1996_SoftwarePatterns,porter_sequences_2005,Zdun2007_FormalPatternLanguages} (a representation we use in figures throughout this paper), we use a graph-based representation to visualize the graph structure of pattern languages and pattern views in the graphical editor.
Within the graphical editor, relations of a pattern can be inspected in detail, and new relations can be added by drawing arrows between two patterns.
Additionally, the visualization of the graph can be adapted by re-positioning nodes, zooming in and out, and triggering an automatic reordering of the graph layout based on the edges.
Users can therefore directly edit or interact with the visualized pattern graph and observe how new relations or patterns lead to structural changes as the overall structure of the \emph{network of patterns} can be grasped immediately.
In the course of this work we not only conceptually extended the pattern repository but also refactored the implementation.
The user interface of the pattern repository was reimplemented as an Angular frontend\footnote{https://github.com/PatternPedia/pattern-pedia-views-ui} and we used Spring Boot for implementing a RESTful backend\footnote{https://github.com/PatternPedia/pattern-pedia-views-backend}.

\section{Case Study}\label{sec:use-case}

In our motivating scenario in \cref{subsec:CrossLanguageRelations}, we stated that patterns from multiple domains are needed for realizing a secure elastic cloud application. 
In this section, we present a case study with the pattern view for the context of \emph{secure elastic cloud applications}.
\begin{figure}[!t]
    \centering
    \includegraphics[width=0.7\textwidth]{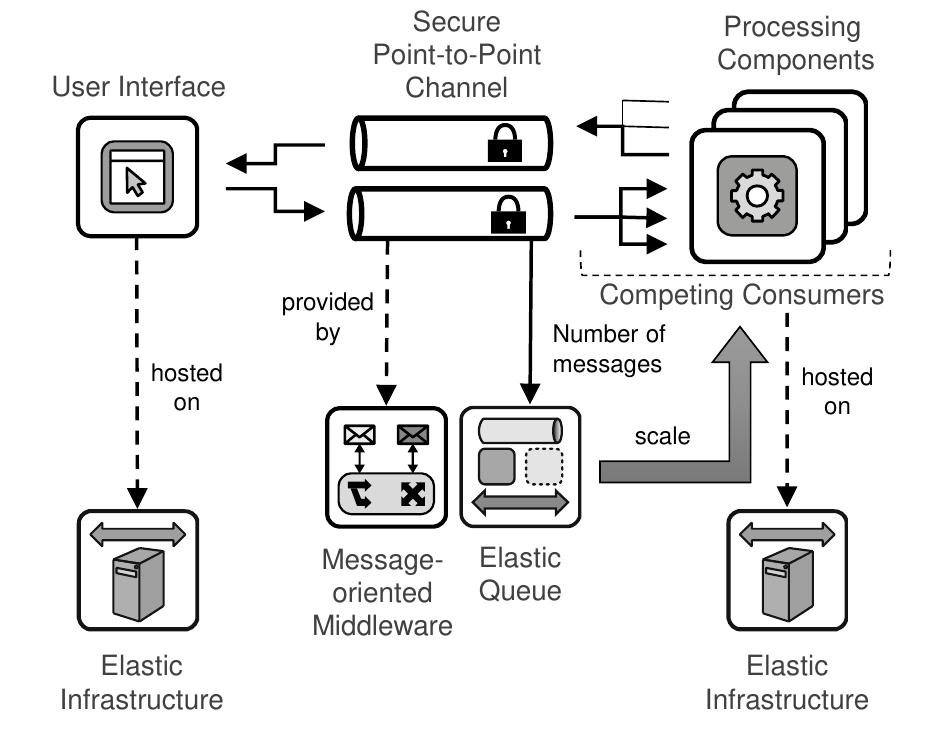}
    \caption{Architecture of a secure elastic cloud application.}
    \label{fig:usecase_architecture}
\end{figure}

Users expect certain applications to be always available.
To fulfil these expectations, cloud providers offer infrastructure and services that can be used to guarantee the availability of an application even for a sudden increase of demand. 
Elastic cloud application deal with changing demand by adjusting the amount of resources that are used for the application~\cite{Fehling2014_Book}.
In addition, data security plays a major role, especially when data is exchanged between communication partners.

\cref{fig:usecase_architecture} depicts the architecture of a secure elastic cloud application.
The application consists of a  \emph{User Interface} component that communicates with \emph{Processing Components} via messaging.
Both components are hosted on an \emph{Elastic Infrastructure}.
The number of messages in the channel is monitored to determine the workload of the processing component instances. 
Depending on the number of messages, the \emph{Elastic Queue} adjusts the number of instances.
As any processing component instance can answer a request, the component is implemented as \emph{Stateless Component} and its instances act as \emph{Competing Consumers} listening on a \emph{Point-to-Point} channel provided by a \emph{Message-Oriented Middleware}.
After consuming and processing a message the processing component instance can send an answer via another point-to-point channel. 
To ensure data security, the communication between the component must be encrypted.

\begin{figure*}[t]
    \centering
    \includegraphics[width=\textwidth]{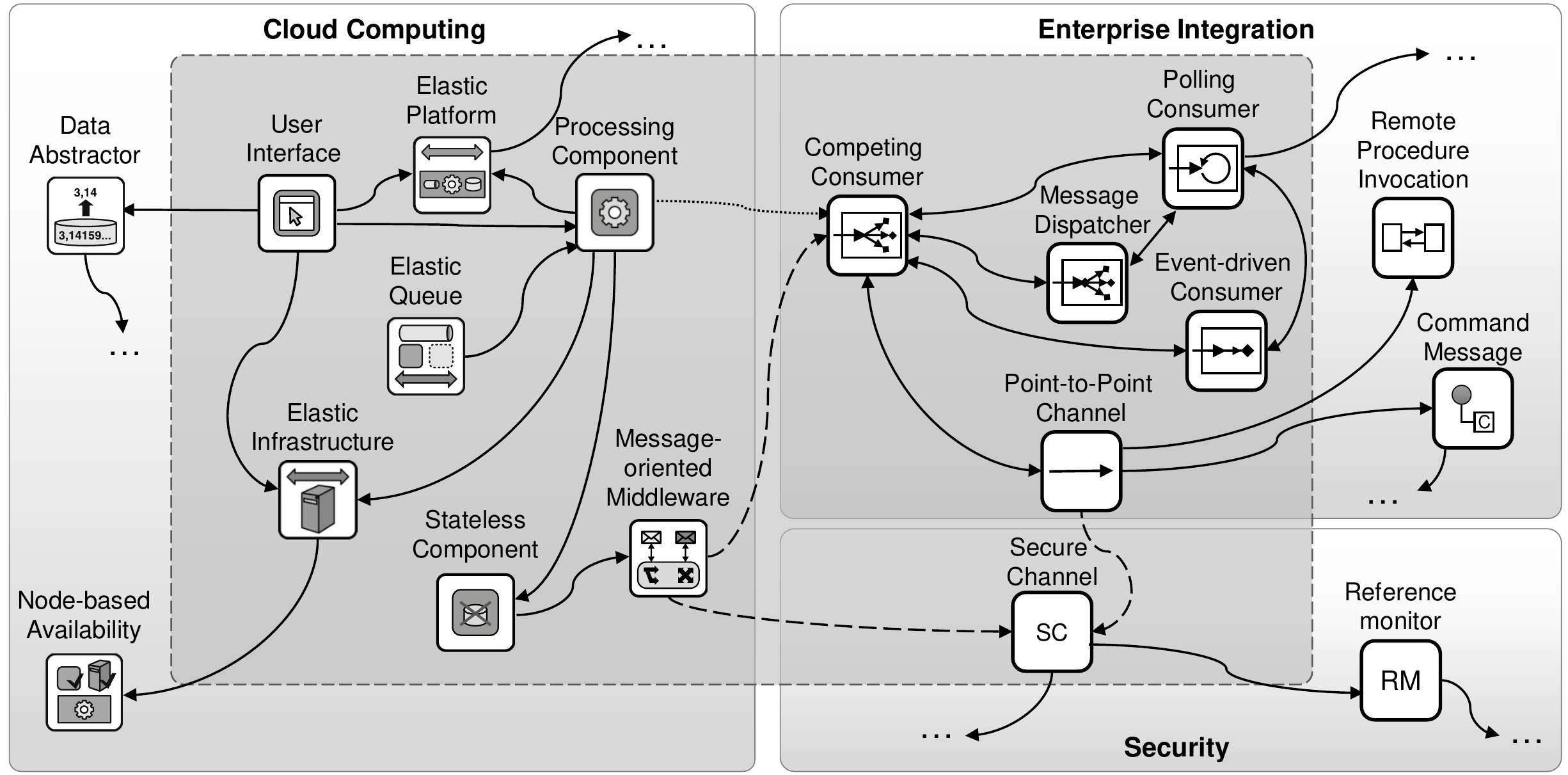}
    \caption{Pattern view for secure elastic cloud applications, 
    \\refer to \cref{fig:motivation} for a legend of the relation types.}
    \label{fig:elastic_view}
\end{figure*}

For such an application, there are a number of patterns that should be taken into account during implementation.
In \cref{fig:elastic_view}, the pattern view for secure elastic cloud applications is shown.
It includes patterns from the cloud computing, enterprise integration, and security pattern languages which are relevant in this specific context.
Besides existing relations from the original pattern languages, three new cross-language relations (dashed arrows) are contained in the pattern view. 
In addition to the already named patterns also a \emph{Message Dispatcher} can be used to delegate the message to one specific consumer, i.e. one processing component instance.
Each \emph{Competing Consumer} can be implemented as \emph{Polling Consumer}, \emph{Event-Driven Consumer}, or a combination of both~\cite{Hohpe2003_PatternLanguages}.
A \emph{Message-oriented Middleware} provides the functionality for communication via messaging and therefore also the message channels for competing consumers.
To ensure that a message is consumed only once, the consumers must all listen to the same \emph{Point-to-Point Channel}.
As all transferred data of the application must be encrypted, the point-to-point channel must also implement the \emph{Secure Channel} pattern.
Once defined, this pattern view can be used by other cloud application architects to realize their secure elastic cloud applications. 
Since the existing knowledge is only enriched by the pattern views, further relevant patterns outside the view can be identified by the existing relations in the pattern languages.


\section{Related Work}\label{sec:related-work}
Several authors have examined relations and patterns across multiple pattern languages. 
Avgeriou~\&~Zdun~\cite{Avgeriou2005_Architecturalpatterns} reviewed architectural views and patterns from different languages. 
They assigned each architectural pattern to its primary architectural view and defined relations between the patterns.
As each of their collection of patterns and relations for an architectural view is worth documenting, we adopted the idea of views as a concept that is not limited to the domain of IT architecture. 
Caiza~et~al.~\cite{caiza_2017_organizing} standardize the relations of various privacy pattern languages to combine them into a new pattern language.
Porter~et~al.~\cite{porter_sequences_2005} derived a method to combine pattern languages based on pattern sequences. 
In contrast, pattern views contain only those patterns of different languages and their relations that are relevant in a certain context. Thus, pattern views are more specific and less complex than potentially large combined pattern languages. 

Buschmann~et~al.~\cite{Buschmann1996_POSA1} introduce pattern stories as textural descriptions that walk a reader through the application of multiple patterns. 
In an exemplary pattern story, they demonstrate how patterns of multiple languages are used together. However, pattern stories are targeted at illustrating common pattern sequences.
Pattern views are not limited to express sequential steps but can express arbitrary relationships. 

Reinfurt~et~al.~\cite{reinfurt_where_2019} present an algorithm for finding entry points in pattern languages. 
Their algorithm can be used to support several manual steps that are needed to document pattern views: 
For a formalized set of problems related to the context of the pattern view, their algorithm suggests suitable languages and a pattern that serves as a starting point.

Köppe~et~al.~\cite{Koeppe2016_TowardsExtendingOnlinePatternRepositories} elaborated requirements for online pattern repositories. 
He used the term pattern views in the sense that there should be different options (pattern views) for displaying a pattern, e.g., for smaller screens or optimized for printing. 
His notion of a pattern view, therefore, defines the visual representation of a pattern whereas we use the term pattern view for a concept to encompass patterns and relations that are relevant for a particular context.
Apparently similar terms from other domains are \emph{process views} and \emph{process viewing patterns}~\cite{Schumm2010_ProcessViewingPatterns}. 
Process views are used to represent a complex business process regarding certain aspects, e.g. by reducing the process to its essentials parts~\cite{Schumm2010_ProcessViewingPatterns}. 
They are obtained by applying transformations on the process graph of the business process~\cite{Schumm_2010_processViewsCompliance}. 
These transformations have been described by process viewing patterns~\cite{Schumm2010_ProcessViewingPatterns}. 
In contrast to pattern views that are created by selecting suitable nodes (patterns) and redefine the relations (edges) between them, these transformations can be far more sophisticated, e.g., nodes of an process graph can be aggregated. 

Pavli\v{c}~et~al~\cite{pavlic_improving_2008} introduced the concept of pattern-containers to represent pattern collections.  
They formalized how patterns represent knowledge in an ontology.
Relations are modeled by specifying that a pattern is related to another pattern. 
Unfortunately, in their ontology, the relation can not be described further and thus, the type of the relation can not be defined.
They define pattern-containers as a way to create pattern collections:
Pattern-containers can include patterns and other pattern-containers. 
A pattern can be included in multiple pattern-containers.
But given their ontology, pattern-containers can not be used to represent pattern views: 
As it can not be defined what relations are relevant for a pattern-container, they represent a simple subset of patterns. 

Graph-based representations for pattern languages are commonly used to reflect Alexander's description of a network of patterns~\cite{Buschmann1996_POSA1,Falkenthal2018_NatureOfPatternLanguages,Zdun2007_FormalPatternLanguages}. 
Another pattern repository, \emph{The Public Sphere Project} mentions that a graph representation of all of their patterns and relations was
once created~\cite{publicSphereProject_repo}, but only a sub-graph of it (8 patterns and their
relations) can still be found on their website. 
Nevertheless, even the complete graph is still a static representation of their underlying living pattern network. 
Schauer~\&~Keller~\cite{schauer_pattern_Visualization} developed a tool for documenting software systems.
Although they use patterns to enhance multiple graph-based views for a software system (e.g. as annotations in UML diagrams), they do not offer a general view on patterns. 
Welicki~et~al.~\cite{welicki_model_2005} developed a visualization tool that can be used to search and create relations (including cross-language relations) between patterns in their software pattern catalog.
They also implemented different views on a pattern that display e.g. a summary or software-specific views (e.g. source-code of a concrete implementation).
The MUSE repository of Barzen~\cite{barzen_2018_disseration} offers a graph-based representation of concrete costumes that occur in films and are understood as concrete solutions for costume patterns.
However, these tools and repositories do not offer different perspectives on the relations of the patterns or pattern languages.
Therefore, no other pattern repository or tool known to us offers graph-based representations of pattern languages and the ability to ~\cite{barzen_2018_disseration} dynamically combine patterns from different languages to pattern views for a particular problem context.

\section{Conclusion and Future Work}
\label{sec:conclusion}
In this paper, we introduced the concept of pattern views to explicitly document cross-domain knowledge relevant for a particular problem context.
Patterns from either different pattern languages or one pattern languages that are relevant for a specific problem can be combined to a so-called pattern view.
In addition to the existing patterns and relations of the underlying pattern languages, view-specific relations can be defined if necessary for the given context.
Therefore, cross-domain knowledge expressed by these relations is documented explicitly and within a meaningful context.

We extended our pattern repository that was presented in previous works~\cite{Fehling2014_Paper_PatternPedia,Falkenthal_2017_SolutionLanguages} by the concept of pattern views. 
Therefore, our repository allows to collect multiple pattern languages, and to define pattern views that can combine, reuse, and extend the structure of pattern languages that is given by the patterns and their relations.
Our repository also offers a graph-based representation for the visualization of pattern views and pattern languages that visualizes the \emph{network of patterns}. 
We plan to collect further pattern languages in the repository, such as Internet of Things patterns~\cite{Reinfurt_2017_iot_patterns_devices} or green IT patterns~\cite{Nowak2011_GreenBusinessProcessPatterns} and to extend our collection of pattern views. We will further evaluate if some pattern needs to be adapted to be used in the context of a pattern view.
For future research we will especially consider patterns from new research areas such as music~\cite{barzen_2016_muse4music} or Quantum Computing~\cite{Leymann2019_Patterns_QuantumAlgorithms}. Patterns for quantum computing are interesting as new technologies need to be integrated into our current software systems (for which we already have patterns at hand). 
Also, an open access hosting of the pattern repository would offers multiple advantages in future.


\section*{Acknowledgment}
This work was partially funded by the BMWi projects \textit{PlanQK (01MK20005N)} and \textit{IC4F (01MA17008G)}. The authors would like to thank Lisa Podszun for her help with the documentation of existing patterns.

\bibliographystyle{IEEEtran}
\bibliography{patternview-paper}

\end{document}